\documentclass[12pt]{article}
\usepackage[centertags]{amsmath}
\usepackage{amsfonts}
\usepackage{amssymb}
\usepackage{amsthm} 
\usepackage{newlfont}
\usepackage{epsfig}
\usepackage{amscd}
\newcommand{\beq}{\begin{equation}}
\newcommand{\eeq}{\end{equation}}
\newcommand{\ba}{\begin{array}}
\newcommand{\ea}{\end{array}}
\newcommand{\bea}{\begin{eqnarray}}
\newcommand{\eea}{\end{eqnarray}}

\begin{document}

\begin{center}
{\large \sc \bf Particular solutions to  multidimensional PDEs represented 
in the form of one-dimensional flow
}

\vskip 15pt

{\large  A. I. Zenchuk }

\vskip 8pt

\smallskip

{\it  Institute of Chemical Physics, RAS,
Acad. Semenov av., 1
Chernogolovka,
Moscow region
142432,
Russia}

\smallskip

\vskip 5pt

e-mail:  {\tt zenchuk@itp.ac.ru }

\vskip 5pt

{\today}

\end{center}

\begin{abstract}
We represent an algorithm  reducing 
the $(M+1)$-dimensional  
nonlinear partial differential equation (PDE)  representable in the form of one-dimensional flow 
$u_t + w_{x_1}(u,u_{
x},u_{
xx},\dots )=0$,
(where $w$ is an arbitrary local  function of $u$ and its $x_i$-derivatives, 
$i=1,\dots,M$) to the  family of  $M$-dimensional nonlinear PDEs $F(u,w)=0$, where $F$ is 
general (or particular) solution of a certain   second order two-dimensional 
nonlinear PDE. 
Particularly, the $M$-dimensional  PDE might be an ODE which, in some cases, may be  
integrated yielding the explicite solutions to the original ($M+1$)-dimensional  PDE.  
Moreover, the spectral parameter may be introduced into the function $F$
which
yields a linear spectral equation associated with the original  PDE.
Simplest examples of nonlinear PDEs with explicite solutions are given. 
\end{abstract}

%%%%%%%%%%%%%%%%
\section{Introduction}

It is well known that the method of characteristics 
\cite{Whitham} allows one to integrate  the  first order 
nonlinear  partial differential equations (PDEs) in arbitrary dimensions. This
approach seemed out 
to be very useful in study of the (1+1)-dimensional systems of hydrodynamic type, where 
a modification of this method (the holograph method)  was developed \cite{T1,DN,T2,F}.  
However, introduction of the higher order derivatives into the nonlinear PDE
either 
destroys the integrability or 
requires a different
 integration algorithm.  For instance, the (1+1)-dimensional inviscid B\"urgers equation is
 the simplest nonlinear PDE completely
integrable by the 
method of characteristics:
\begin{eqnarray}\label{br}
u_t -  u u_{x} =0. 
\end{eqnarray}
Adding the second order derivative into this equation we obtain the 
 well known viscous B\"urgers equation:
\begin{eqnarray}\label{br00}
u_t+u_{xx} - u u_{x} =0.
\end{eqnarray}
This equation is also completely integrable, although a different tool must be applied.
Namely, it is linearizable   by the Hopf-Cole substitution \cite{H,C,Calogero}.
However, the higher-dimensional viscous B\"urgers system may not be  completely integrated 
through  linearization  and must be studied using a different approach. For
instance, 
its solutions with the finite time blow up were studied 
by the renormalization group method in refs.\cite{LS1,LS2}.
Next, introducing the third order derivative into eq.(\ref{br})  we obtain the
Korteweg-de Vries equation (KdV)
\begin{eqnarray}
u_t+u_{xxx} - u u_{x}=0
\end{eqnarray}
integrable by  the inverse spectral transform method (ISTM)
\cite{GGKM,ZMNP,AC,Konopelchenko,ZSh1,ZSh2} which is another productive tool for
integration of 
a large class of nonlinear PDEs. 

It is shown in ref.\cite{Z_KdV}  that, deforming the characteristics of the
(1+1)-dimensional 
PDE (\ref{br}),
we reveal a family of particular solutions to a large  class of 
multidimensional higher order nonlinear evolutionary 
PDEs with the KdV-type nonlinearity described by  $(M-1)$- (or $M-2$) 
dimensional nonlinear PDEs. In particular, there is a class of (2+1)-dimensional
PDEs 
with solutions satisfying the appropriate ODE, which, in turn, might have
explicite solutions. 

In this paper we extend  the algorithm of ref.\cite{Z_KdV} 
 to the PDEs writable in the form of one-dimensional flow 
\begin{eqnarray}\label{intr}\label{uw}\label{ex2ut1}
u_t + w_{x_1}(u,u_{ x},u_{{xx}},\dots )=0,
\end{eqnarray}
where $w$ is an arbitrary local function of $u$ and its $x_i$-derivatives,
$i=1,\dots,M$,  $M$ is the number of variables $x_i$ in the list
$x=\{x_1,x_2,\dots, x_M\}$. 
We show that this equation possesses a large manifold of solutions described by 
the equation $F(u,w)=0$ (or $F(x_1+ u t,w)=0$) where $F$ satisfies
a certain 
two-dimensional second order PDE. This PDE  holds for any 
nonlinear 
PDE representable in the form (\ref{intr}). 
Simplest examples of PDEs together with   solutions are represented. 

The structure of this paper is following. In Sec.\ref{Section:first},
we consider the nonlinear PDE (\ref{intr}) with 
$w=-\frac{1}{2} u^2 + f(u_{ x},u_{{xx}},\dots ) $, whose particular 
solutions involve the characteristics of 
eq.(\ref{br}). Eq.(\ref{intr}) with arbitrary $w$ is considered in 
Sec.\ref{Section:second}. In this case solution of PDE does not depend 
on the characteristics of eq.(\ref{br}). 
The second order PDE for the function $F$ is discussed in 
Sec.\ref{Section:solutionsF}. The spectral problem for the nonlinear PDE (4)
is derived in Sec.\ref{Section:spectral}, where the  richness of the solution space 
is discussed as well. Particular cases of solvable  
equation $F=0$  are represented in Sec.\ref{Section:F}. 
Conclusions are given in Sec.\ref{Section:conclusions}.

%%%%%%
\section{Equation $u_t - u u_{x_1}  + f_{x_1}(u_{ x},u_{{xx}},\dots) =0$} 
\label{Section:first}

In this section we propose an algorithm for construction of a family of $M$-dimensional reductions  to the
nonlinear PDE writable in the form (\ref{uw})
with $w$ having a particular form
\begin{eqnarray}\label{w}
w=  -\frac{1}{2} u^2 + f(u_{ x},u_{{xx}},\dots),
 \end{eqnarray}
 which generalizes the equations considered
in  ref.\cite{Z_KdV}.
Here $f$ is an arbitrary function of any order derivatives of $u$ 
with respect to $x_i$, $i=1,\dots, M$, but independent on $u$.
Since the total $x_1$-derivative $w_{x_1}$ may be written as
\begin{eqnarray}\label{wQu}
w_{x_1} = Q u_{x_1},
\end{eqnarray}
where $Q$ is the differential operator
\begin{eqnarray}\label{Q}
Q=-u +\sum_{i=1}^M f_{ u_{x_i}} \partial_{ x_i} +
 \sum_{i_1,i_2=1}^M f_{ u_{x_{i_1}x_{i_2}}} \partial_{ x_{i_1}x_{i_2}} + 
 \dots,
\end{eqnarray}
we may rewrite eq.(\ref{intr}) as
\begin{eqnarray}\label{ex1ut20}\label{ex1ut2}
E(u)=u_t + Qu_{x_1} =0.
 \end{eqnarray}

It is important that the function $w$ satisfies  the linear equation
\begin{eqnarray}\label{ex1ut3}
E(w)=w_t +Qw_{x_1}  =0,
\end{eqnarray}
which may be shown directly.
Further we need the following evident relation between $E(u)$ and $E(w)$:
\begin{eqnarray}\label{uw2}
E(w) = Q E(u),
\end{eqnarray}
which may be derived applying the operator $Q$ to eq.(\ref{ex1ut2}) and using
relation 
(\ref{wQu}) together with $Qu_t=w_t$.
The family of  $M$-dimensional reductions of eq.(\ref{intr}) with $w$ defined in 
eq.(\ref{w}) may be constructed 
using the following theorem.

{\bf Theorem 1.}
Let the function $u$ be a solution of the following nonlinear $M$-dimensional 
PDE:
\begin{eqnarray}\label{ex1F}
F(\xi, w)=0,\;\;\xi=x_1+ t u,
\end{eqnarray}
where the function $F(\xi,w)$ satisfies the two-dimensional second order PDE:
\begin{eqnarray}\label{F}
F_{\xi\xi} + F_{ww} \frac{F_\xi^2}{F_w^2} -2 F_{\xi w} \frac{F_\xi}{F_w} =M(F,\xi,w),\;\;\xi=t u + x_1 ,
\;\;F_w|_{F=0}\neq 0.
\end{eqnarray}
Here $M$ is some function of $F$, $\xi$ and $w$  satisfying the condition 
\begin{eqnarray}\label{MF0}
\label{M1}
M(F,\xi,w)|_{F=0} =0.
\end{eqnarray}
Then the function $\psi$, 
\begin{eqnarray}\label{psi}
\psi\equiv E(u) = u_t +w_{x_1},
\end{eqnarray}
is a solution of  the linear PDE
\begin{eqnarray}\label{Fpsi}
\Big(F_{\xi} t + F_{w} Q\Big) \psi =0.
\end{eqnarray}
If, in addition, $\psi$ satisfies the 
 zero initial-boundary conditions, then $u$ is a solution of  nonlinear PDE  
(\ref{intr}) with $w$ given in form (\ref{w}).

{\it Proof.}
To derive the nonlinear PDE  (\ref{intr}) from eq.(\ref{ex1F}), we first 
differentiate eq.(\ref{ex1F})  with respect to $x_i$ and $t$:
\begin{eqnarray}\label{x1}
&&
E_{x_1}:=F_\xi (1+t u_{x_1}) + F_w w_{x_1} =0,\\\label{xi}
&&
E_{x_i}:=F_\xi (t u_{x_i}) + F_w w_{x_i} =0,\;\;i>1,\\\label{t}
&&
E_t:=F_\xi (u+t u_{t}) + F_w w_{t} =0.
\end{eqnarray}
If $F_\xi|_{F=0}=F_w|_{F=0}=0$, then eqs.(\ref{x1}-\ref{t}) become identities.
We  assume that  $F_w|_{F=0}\neq 0$ (the case $ F_\xi|_{F=0}\neq 0$ can be
treated similarly).  
Then solving eqs. (\ref{x1}) and (\ref{xi}) with respect to $w_{x_1}$ and
$w_{x_i}$ ($i>1$) respectively  we obtain:
\begin{eqnarray}\label{wx}\label{wx2}
&&
w_{x_i}= -\frac{F_\xi}{F_w}  (\delta_{1i} +t u_{x_i}),\;\;i\ge 1.
\end{eqnarray}
Let us consider the second derivatives of eq.(\ref{ex1F}) with respect to an
arbitrary 
pair $x_i$ and $x_j$ and substitute eqs.(\ref{wx}) for the  derivatives 
$w_{x_i}$, 
$i\ge 1$:
\begin{eqnarray}\label{F0x}
F_{\xi} t u_{x_ix_j} + F_w w_{x_ix_j} +
(\delta_{1i}+t u_{x_i})(\delta_{1j} +t u_{x_j} )\left(F_{\xi\xi} + F_{ww}
\frac{F_\xi^2}{F_w^2} -2 F_{\xi w} \frac{F_\xi}{F_w} \right)  =0.
\end{eqnarray}
If  $F$ satisfies the  nonlinear PDE (\ref{F}), then the last term in
eq.(\ref{F0x}) 
vanishes reducing  eq.(\ref{F0x}) to
\begin{eqnarray}\label{F01}
F_{\xi} t u_{x_ix_j} + F_w w_{x_ix_j}  =0.
\end{eqnarray}
Solving this equation with respect to $w_{x_ix_j}$, we obtain the relation
between 
the second derivatives of the functions $w$ and $u$:
\begin{eqnarray}\label{wxx}
w_{x_ix_j} =-\frac{F_{\xi}}{F_w} t u_{x_ix_j} .
\end{eqnarray}
By induction, for  any order derivative
$D=\prod_{i=1}^M\partial_{x_i}^{n_i}$ (where $n_i$ are arbitrary integers),   
we obtain the following relation:
\begin{eqnarray}\label{D}
F_{\xi} t D u + F_w  D w  =0.
\end{eqnarray}
Consequently, for the differential operator $\tilde Q = Q+u$ (with the operator
$Q$ 
given by expression (\ref{Q})) we may write
\begin{eqnarray}\label{F01g}
E_{\tilde Q}:=F_{\xi} t \tilde Q u_{x_1} + F_w  \tilde Q w_{x_1}  =0.
\end{eqnarray}
Now we consider the following combination of
eqs.(\ref{x1},\ref{t},\ref{F01g}): 
\begin{eqnarray}
\label{xtL}
E_t  -  u  E_{x_1} +E_{\tilde Q} =0,
\end{eqnarray}
which reads
\begin{eqnarray}\label{Fuw}
F_{\xi} t E(u)  + F_{w} E(w) =0. 
\end{eqnarray}
In virtue of relation (\ref{uw2}), we may write 
eq.(\ref{Fuw}) as eq.(\ref{Fpsi}) with $\psi$ given by expression (\ref{psi}).
If  $\psi$ 
satisfies the zero initial-boundary conditions, then $\psi\equiv E(u) \equiv 0$, which is
equivalent to eq.(\ref{intr},\ref{w}). $\Box$

Now we shall give several remarks.
\begin{enumerate}
\item
Domain of variables $x_i$, $i=1,\dots,M$, might be either bounded or unbounded. 
\item
For the particular choice of $F$,
\begin{eqnarray}\label{partF}
F(\xi,w)= \xi+w,
\end{eqnarray}
eq.(\ref{ex1F}) transforms to the form considered in \cite{Z_KdV} 
\item
If the function $u$ is a solution of  eq.(\ref{uw},\ref{w}) and of eq.(\ref{ex1F}) with
$F_w|_{F=0}\neq 0$, 
then it satisfies the equation
\begin{eqnarray}
\label{uwF}
u_t - \frac{F_\xi}{F_w} ( 1+ t u_{x_1}) =0.
\end{eqnarray} 
In particular, if $F$ is taken in the form (\ref{partF}), then eq.(\ref{uwF})
reduces to 
\begin{eqnarray}\label{uwFred}
u_t=t u_{x_1} +1,
\end{eqnarray} 
obtained in \cite{Z_KdV}.
To derive eq.(\ref{uwF}) we differentiate eq.(\ref{ex1F}) with respect to $x_1$,
solve the resulting equation for
$w_{x_1}$ and substitute it into eq.(\ref{uw}).
\item
Since the differential
operator $Q$ (\ref{Q})
 is defined by the function $f$ in expression for $w$ (\ref{w}), then
 the zero boundary conditions imposed on the function $\psi$
(\ref{psi}) 
are completely defined by the mentioned above function $f$  and does not
depend on 
the particular function $F(\xi,w)$.  
\end{enumerate}

%%%%%%%%%%%
\subsection{Example}
\label{Section:Example1}
As an example we consider the nonlinear PDE 
\begin{eqnarray}
u_t + u_{x_1x_2} - u u_{x_1} +\alpha (u_{x_2}^2)_{x_1}=0,
\end{eqnarray}
which may be viewed as a deformation of equation 
\begin{eqnarray}\label{ex10}
u_t + u_{x_1x_2} - u u_{x_1}=0
\end{eqnarray}
 considered in ref.\cite{Z_KdV}.
In this case
\begin{eqnarray}
w=u_{x_2} - \frac{u^2}{2} + \alpha u_{x_2}^2 .
\end{eqnarray}
Using eq.(\ref{partF}) as a  simple solution of eq.(\ref{F}) with $M=0$,
we write eq.(\ref{ex1F}) as
\begin{eqnarray}\label{exF1}
u_{x_2} + \alpha u_{x_2}^2  = \frac{u^2}{2}-t u - x_1.
\end{eqnarray}
Solving this equation for $u_{x_2}$ we obtain
\begin{eqnarray}\label{u}
u_{x_2} =-\frac{1}{2\alpha} \left(1\pm 
\sqrt{1+2 \alpha (u^2 - 2 (t u + x_1) ) }
\right).
\end{eqnarray}
Integration of this ODE yields a solution $u$ in the following  implicit form:
\begin{eqnarray}\label{hhh}
h_{\pm}(u,x_1,t) = C(x_1,t)+ x_2,
\end{eqnarray}
where
\begin{eqnarray}\label{hh}
h_{\pm}(u,x_1,t)&=&\frac{1}{2\sqrt{ \eta }}
\left(\ln\frac{\sqrt{\eta}-u+t}{\sqrt{\eta}+u-t}
\pm\ln\frac{\sqrt{\eta}-u+t}{\sqrt{\eta}+u-t}
\pm\right.\\\nonumber
&&
\left.
\ln\frac{2 \alpha (\eta +\sqrt{\eta} (u-t))-\sqrt{2 \alpha (u-t)^2 - 2 \alpha
\eta +1}-1 }{
2 \alpha (\eta -\sqrt{\eta} (u-t))-\sqrt{2 \alpha (u-t)^2 - 2 \alpha \eta
+1}-1}\right)
\pm\\\nonumber
&&\sqrt{2\alpha}\ln(\sqrt{2 \alpha} (u-t) +\sqrt{2 \alpha (u-t)^2 - 2 \alpha
\eta +1}),
\end{eqnarray}
with
\begin{eqnarray}\label{eta}
\eta= t^2 + 2 x_1.
\end{eqnarray}
Here the function $C$ may not be arbitrary but must provide the
zero initial-boundary condition for the function $\psi$,
\begin{eqnarray}\label{psi1}
\psi= u_t + u_{x_1x_2} - u u_{x_1} +\alpha (u_{x_2}^2)_{x_1},
\end{eqnarray}
considered as a  solution to eq.(\ref{Fpsi})  with $F_\xi=F_w=1$ and the 
first order one-dimensional  differential operator 
 $Q=((1+2 \alpha u_{x_2})\partial_{x_2} - u)$. Thus we need a single boundary
condition  at the boundary 
point, say $x_2=0$:
\begin{eqnarray}\label{psiex1}
\psi|_{x_2=0}=0 \;\;{\mbox{with}} \;\; u|_{x_2=0} =\chi(x_1,t).
\end{eqnarray}
Writing eq.(\ref{uwFred}) at the boundary point $x_2=0$ we obtain
\begin{eqnarray}
\chi_t = t \chi_{x_1} +1.
\end{eqnarray}
Thus
\begin{eqnarray}\label{chi}
\chi= A(\eta) + t  .
\end{eqnarray}
Substituting  $\chi$ from eq.(\ref{chi})  into eqs.(\ref{hhh},\ref{hh}) with $x_2=0$ we
conclude that 
$C$ is the function of $\eta$ only. This, in tern, 
allows us to consider $C(\eta)$ as an arbitrary function of $\eta$ in
eq.(\ref{hhh}), 
while the boundary condition (\ref{psiex1})  may be taken as an implicit
definition of
$A(\eta)$ in terms of the function $C(\eta)$:
\begin{eqnarray}\label{C1}
C(\eta)&=& h_{\pm}(A(\eta) + t,x_1,t)\;\;\Rightarrow
\\\nonumber
C(\eta)&=&\frac{1}{2\sqrt{ \eta }}
\left(\ln\frac{\sqrt{\eta}-A(\eta)}{\sqrt{\eta}+A(\eta)}
\pm\ln\frac{\sqrt{\eta}-A(\eta)}{\sqrt{\eta}+A(\eta)}
\pm\right.\\\nonumber
&&
\left.
\ln\frac{2 \alpha (\eta +\sqrt{\eta} A(\eta))-\sqrt{2 \alpha A(\eta)^2 - 2
\alpha \eta +1}-1 }{
2 \alpha (\eta -\sqrt{\eta} A(\eta))-\sqrt{2 \alpha A\eta^2 - 2 \alpha \eta
+1}-1}\right)
\pm\\\nonumber
&&\sqrt{2\alpha}\ln(\sqrt{2 \alpha} A(\eta) +\sqrt{2 \alpha A\eta^2 - 2 \alpha
\eta +1}).
\end{eqnarray}

If $\alpha<<1$, then expansion of (\ref{hh}) in powers of $\alpha$ 
(up to the linear in $\alpha$ term) yields
\begin{eqnarray}\label{hhser}\label{h+}
&&
h_+=\frac{1}{\sqrt{\eta }}
\ln\frac{\sqrt{\eta}-u+t}{\sqrt{\eta}+u-t} +\sqrt{2\alpha} 
\ln\,2+ \alpha (u-t) + o(\alpha),\\\label{h-}
&&
h_-=- \alpha (u-t) -\sqrt{2\alpha} \ln\,2 + o(\alpha).
\end{eqnarray}
Thus, the solution $u$ corresponding to $h_-$ in eq.(\ref{hhh})  becomes
singular as
$\alpha\to 0$. The solution $u$
corresponding to $h_+$ can be written as 
\begin{eqnarray}
u=u_1+\frac{\alpha}{2}(C_2(\eta) - u_1) (u_1^2-\eta) +o(\alpha)
\end{eqnarray}
where $u_1$ is a solution of eq.(\ref{ex10}) obtained in ref.\cite{Z_KdV}:
\begin{eqnarray}
&&u_1=t+\sqrt{\eta}\frac{1+\hat C(\eta) e^{x_2 \sqrt{\eta}}}{
1-\hat C(\eta) e^{x_2 \sqrt{\eta}}},\\\nonumber
&&
\hat C(\eta)= e^{ (C_1(\eta)-\sqrt{2\alpha} \ln\,2)\sqrt{\eta}},
\end{eqnarray}
and we represent an arbitrary function $C(\eta)$ of eq.(\ref{hhh}) in the form 
$C(\eta)=C_1(\eta)+\alpha C_2(\eta)$.
Considering solutions having finite asymptotics as $t\to\infty$, we note that 
 $u_1\to 0$ as $t\to\infty$.
 Thus, if  $C_2(\eta)=\frac{\tilde C_2(\eta)}{\eta}$ and $\tilde C_2(\eta) 
 \to c_2=const$ as $t\to\infty$, then $u\to -\frac{\alpha c_2}{2}$ as 
$t\to\infty$.
 If $C_2(\eta)\to 0 $ as  $t\to\infty$, then $u\to 0$ as well (up to
$o(\alpha)$).

%%%%%%%%%%%%%%%%%%%%%%%%%%%%%%%%%%%
\section{Equation (\ref{intr}) with arbitrary $w$}
\label{Section:second}
In this section we modify  an algorithm of Sec.\ref{Section:first} to solve
equation of more general form (\ref{intr}),
where $w$ is an arbitrary function of $u$ and  its derivatives  
with respect to $x_i$, $i=1,\dots, M$.
Now we may rewrite eq.(\ref{ex2ut1}) as eq.(\ref{ex1ut2}) 
where  the linear differential operator $Q$ reads:
\begin{eqnarray}\label{QQ}
Q=w_u  + \sum_{i=1}^M w_{ u_{x_i}} \partial_{ x_i} +
 \sum_{i_1,i_2=1}^M w_{ u_{x_{i_1}x_{i_2}}} \partial_{ x_{i_1}x_{i_2}} + \dots.
\end{eqnarray}
It is important that the function $w$ satisfies the linear equation
(\ref{ex1ut3}) with
$Q$ given by (\ref{QQ}).
Relation (\ref{uw2}) holds as well.

The family of $M$-dimensional reductions for eq.(\ref{intr}) may be constructed using the following theorem, 
which is similar to  Theorem 1 of
Sec.\ref{Section:first}.

{\bf Theorem 2.}
Let the function $u$ be a solution of the following nonlinear $M$-dimensional 
PDE:
\begin{eqnarray}\label{F0}
F(u, w)=0,
\end{eqnarray}
where the function $F(u,w)$ satisfies the PDE:
\begin{eqnarray}\label{F2}
F_{uu} + F_{ww} \frac{F_u^2}{F_w^2} -2 F_{u w} \frac{F_u}{F_w} = M(F,u,w),
\;\;F_w|_{F=0}\neq 0
\end{eqnarray}
and $M$ is some function of $F$, $u$ and $w$  satisfying the condition
\begin{eqnarray}\label{MF02}
\label{M2}
M(F,u,w)|_{F=0} =0.
\end{eqnarray}
Then the function $\psi$, 
\begin{eqnarray}\label{psi2}
\psi \equiv E(u)= u_t + w_{x_1},
\end{eqnarray}
is a solution of the linear PDE
\begin{eqnarray}\label{2Fpsi}
\Big(F_{u}  + F_{w} Q\Big) \psi =0.
\end{eqnarray}
If, in addition, $\psi$ satisfies the  zero initial-boundary conditions, then $u$ is a solution of  nonlinear PDE  
(\ref{ex2ut1}).

{\it Proof.}
The proof of this theorem is quite similar to that of Theorem 1.
To derive  nonlinear PDE  (\ref{ex2ut1}) from eq.(\ref{F0}) we  
differentiate eq.(\ref{F0})  with respect to $x_i$ and $t$:
\begin{eqnarray}\label{2xi}
&&
E_{x_i}:=F_u u_{x_i} + F_w w_{x_i} =0,\;\;i>1,\\\label{2t}
&&
E_t:=F_u  u_{t} + F_w w_{t} =0.
\end{eqnarray}
If $F_w|_{F=0}=F_u|_{F=0}=0$, then eqs.(\ref{2xi},\ref{2t}) become identities.
Assume that $F_w|_{F=0}\neq0$ (the case $F_u|_{F=0}\neq0$ can be treated
similarly). Then 
eq.(\ref{2xi}) yields
\begin{eqnarray}\label{2wx}
&&
w_{x_i}= -\frac{F_u}{F_w}   u_{x_i},\;\;i=1,2,\dots.
\end{eqnarray}
Let us consider  the second derivatives with respect to an 
arbitrary pair 
$x_i$ and $x_j$ and substitute eqs.(\ref{2wx}) for $w_{x_i}$, $i\ge 1$:
\begin{eqnarray}\label{2F0}
F_{u}  u_{x_ix_j} + F_w w_{x_ix_j} +
u_{x_i}u_{x_j} \left(F_{uu} + F_{ww} \frac{F_u^2}{F_w^2} -2 F_{u w}
\frac{F_u}{F_w} \right)  =0.
\end{eqnarray}
If $F$ satisfies  the nonlinear PDE (\ref{F2}), then
 eq.(\ref{2F0})
yields
\begin{eqnarray}\label{2F01}
F_{u}  u_{x_ix_j} + F_w w_{x_ix_j}  =0,\;\;\Rightarrow \;\;
w_{x_ix_j}=-\frac{F_{u}}{F_w}   u_{x_ix_j}
.
\end{eqnarray}
By induction, for  any order derivative 
$D=\prod_{i=1}^M
\partial_{x_i}^{n_i}$
(where $n_i$ are arbitrary integers) we obtain the following relation:
\begin{eqnarray}\label{2F01k}
F_{u}  Du + F_w Dw  =0.
\end{eqnarray}
Consequently, for the differential operator $Q$ 
(\ref{QQ}) we have
\begin{eqnarray}\label{2F01g}
E_Q:=F_{u}  Q u_{x_1} + F_w  Qw_{x_1}  =0.
\end{eqnarray}
Now we may consider the following combination of eqs.(\ref{2t},\ref{2F01g}):
\begin{eqnarray}
\label{xtLex2}
E_t  + E_Q  =0,
\end{eqnarray}
which reads
\begin{eqnarray}\label{Fuwex2}
F_{u}  E(u)  + F_{w} E(w) =0. 
\end{eqnarray}
In virtue of relation (\ref{uw2}) we may write 
eq.(\ref{Fuwex2}) as eq.(\ref{2Fpsi}) with $\psi$ given by expression
(\ref{psi2}).
If, in addition,   $\psi$ 
satisfies the zero initial-boundary conditions, then $\psi\equiv E(u) \equiv 0$, which is
equivalent to eq.(\ref{ex2ut1}). $\Box$

Now we shall give several remarks similar to those given in Sec.\ref{Section:first}
\begin{enumerate}
\item
Domain of variables $x_i$, $i=1,\dots,M$, might be either bounded or unbounded. 
\item
Eq.(\ref{F2}) is equivalent to eq.(\ref{F}) up to the
re-notations $\xi\leftrightarrow u$. 
\item
If the function $u$ satisfies eqs.(\ref{ex2ut1}) and (\ref{F0}) with
$F_w|_{F=0}\neq 0$, 
then it satisfies equation
\begin{eqnarray}
\label{2uwF}
u_t - \frac{F_u}{F_w}  u_{x_1} =0.
\end{eqnarray} 
In particular, if $F$ is taken in the form $F=u+w$, then eq.(\ref{uwF}) reduces
to 
\begin{eqnarray}\label{bex2}
u_t= u_{x_1}.
\end{eqnarray} 
To derive eq.(\ref{2uwF}) we differentiate eq.(\ref{F0}) with respect 
to $x_1$, solve the resulting equation for $w_{x_1}$ 
and substitute it into eq.(\ref{ex2ut1}).
\item
Since the differential
operator $Q$ 
(\ref{QQ})
is defined by the  function $w$ in PDE (\ref{intr}), then the 
zero boundary condition imposed on the function $\psi$
(\ref{psi2}) 
is completely defined by the above function  $w$  and does not
depend on 
the particular function $F(u,w)$.
\end{enumerate}

%%%%%%%%%%%%%%%%
\subsection{Example}
\label{Section:Example2}
As an example, we consider the nonlinear 
PDE 
\begin{eqnarray}\label{pde2}
u_t + u_{x_1x_2} - (u^3)_{x_1}=0.
\end{eqnarray}
In this case
\begin{eqnarray}
w=u_{x_2} -  u^3.
\end{eqnarray}
The simple solution to eq.(\ref{F2}) is following ($M=0$):
\begin{eqnarray}\label{simple_ex2}
F(u,w)=w-\gamma_1 u -\gamma_0,
\end{eqnarray}
which corresponds to the reduction $u_t=-\gamma_1 u_{x_1}$ in eq.(\ref{intr}).
Then eq.(\ref{F0}) reads
\begin{eqnarray}\label{F0ex2}
u_{x_2} = u^3 +\gamma_1 u +  \gamma_0.
\end{eqnarray}
Integration of  ODE (\ref{F0ex2}) yields
\begin{eqnarray}\label{uex20}
&&
\frac{1}{(u_1-u_2)(u_1-u_3)}\ln(u-u_1) +
\frac{1}{(u_2-u_1)(u_2-u_3)}\ln(u-u_2) +\\\nonumber
&&
\frac{1}{(u_3-u_1)(u_3-u_2)}\ln(u-u_3) = x_2 +C(x_1,t),
\end{eqnarray}
where $u_1$, $u_2$ and $u_3$ are roots of the polynomial equation
\begin{eqnarray}
\label{roots}
u^3 +\gamma_1 u +  \gamma_0=0.
\end{eqnarray}
In particular, if $\gamma_0=0$, $\gamma_1=-1$, then  integration of
eq.(\ref{F0ex2}) yields
\begin{eqnarray}\label{uex2}
u^2 = \frac{1}{
1+\tilde C(x_1,t) e^{2 x_2}}.
\end{eqnarray}
The functions  $C$ (or $\tilde C$) 
may not be arbitrary but must provide
the zero initial-boundary condition for the function $\psi$,
\begin{eqnarray}\label{psi1ex2}
\psi= u_t + u_{x_1x_2} - (u^3)_{x_1},
\end{eqnarray}
as a solution to the first order linear PDE (\ref{2Fpsi}) with $F_u=-\gamma_1$, $
F_w=1$ and 
$Q=\partial_{x_2}-3u^2 $. Thus  we have a single boundary condition  at the
boundary point, 
say  $x_2=0$:
\begin{eqnarray}\label{psiex2}
\psi|_{x_2=0}=0 \;\;{\mbox{with}} \;\; u|_{x_2=0} =\chi(x_1,t).
\end{eqnarray}
Eq.(\ref{2uwF}) with $F$ given in eq.(\ref{simple_ex2}) at the boundary point $x_2=0$  reads
\begin{eqnarray}
\chi_t +\gamma_1 \chi_{x_1}= 0.
\end{eqnarray}
Thus
\begin{eqnarray}\label{chiex2}
\chi= A(\eta) ,\;\;
\end{eqnarray}
where
\begin{eqnarray}\label{etaex2}
\eta=x_1 - \gamma_1 t.
\end{eqnarray}
Substituting  $\chi$ from eq.(\ref{chiex2})  into eqs.(\ref{uex20}) and
(\ref{uex2})
with $x_2=0$ we see that the function $C$ (or $\tilde C$) must be a
 function of the single variable $\eta$. 
Thus, $C$ and $\tilde C$ may be taken as  arbitrary functions of $\eta$ in
eqs.(\ref{uex20}) and (\ref{uex2}),
while  boundary condition (\ref{psiex2}) should be considered as a definition
of
the function $A(\eta)$ in terms of $C(\eta)$ or $\tilde C(\eta)$:
\begin{eqnarray}\label{C1ex20}
&&
C(\eta)=\frac{1}{(u_1-u_2)(u_1-u_3)}\ln(A(\eta)-u_1) +
\frac{1}{(u_2-u_2)(u_2-u_3)}\ln(A(\eta)-u_2) +\\\nonumber
&&
\frac{1}{(u_3-u_1)(u_3-u_2)}\ln(A(\eta)-u_3) .
\end{eqnarray}
and
\begin{eqnarray}\label{C1ex2}
A^2(\eta) = \frac{1}{
1+\tilde C(\eta) e^{2 x_2}}.
\end{eqnarray}
If $\tilde C$ is a bounded function of argument, then solution (\ref{uex2}) 
is a bounded solution to eq.(\ref{pde2}).

%%%%%%%%%%%%5

\section{On solutions to nonlinear PDEs  (\ref{F}) and (\ref{F2})}
\label{Section:solutionsF}

Thus we reduce  $(M+1)$-dimensional PDE (\ref{intr}) to $M$-dimensional PDE 
(\ref{ex1F})  (if $w$ is in the 
 form (\ref{w})) or (\ref{F0}) (which holds for any $w$), where 
$F$ satisfies the  two-dimensional second order PDE, respectively, 
eq.(\ref{F}) or 
(\ref{F2}). 
Equations (\ref{ex1F},\ref{F}) are  equivalent  to eqs.(\ref{F0},\ref{F2}) up to
the replacement $\xi\leftrightarrow u$.
We  combine them in the following pair of equations:
\begin{eqnarray}\label{Fpq}
F(p,q)=0,
\end{eqnarray}
and
\begin{eqnarray}\label{gF}
F_{pp} F_q^2 + F_{qq}  F_p^2 -2 F_{pq} F_p F_q ={\cal{M}}(F,p,q),
\end{eqnarray}
which is symmetrical in the variables $p$ and $q$. The variable $p$
must be taken for either $\xi$ or $u$, while the  variable $q$ 
must be taken for 
the variable $w$.
In addition, the function ${\cal{M}}$ must satisfy the  conditions
\begin{eqnarray}\label{calM}
{\cal{M}}(F,p,q)|_{F(p,q)=0} =0.
\end{eqnarray}
Eq.(\ref{gF}) may be viewed as the compatibility condition of $(M+1)$-dimensional PDE (\ref{intr})
and $M$-dimensional PDE (\ref{Fpq}).
In Secs.\ref{Section:first} and \ref{Section:second} we consider examples 
of particular solutions to the nonlinear PDE (\ref{intr}) 
associated with the simplest solution to eq.(\ref{gF}). However, this equation
possesses 
a rich manifold of solutions parametrized by two arbitrary functions of single
variable
 and
deserves the detailed study which is not represented 
in this paper. Below we consider only three particular solutions 
of eq.(\ref{gF})  in the form of a degenerate function of $p$ and $q$: 
\begin{eqnarray}\label{Fdeg}
F(p,q) =\sum_i f_i(p) g_i(q).
\end{eqnarray}

{\bf Example 1.} A simplest solution reads:
\begin{eqnarray}\label{Fex1}
&&
F= \alpha_0+\alpha_1 p + \alpha_2 q ,
\end{eqnarray}
where $\alpha_i$, $i=0,1,2$, are arbitrary constant parameters. 
Namely this solution is used in examples of 
Secs.\ref{Section:Example1} ($\alpha_0=0$, $\alpha_1=\alpha_2=1$) and 
\ref{Section:Example2} ($\alpha_0=-\gamma_0$, $\alpha_1=-\gamma_1$,
$\alpha_2=1$). 
In this case ${\cal{M}}=0$ in eq.(\ref{gF}).

It is worthwhile to note that the local linear expansion of function $F(p,q)$ 
in the neighborhood of a fixed point
$(p_0,q_0)$ reads (we assume that the first  derivatives of $F$ exist in that
point) 
\begin{eqnarray}
F(p,q)\approx F(p_0,q_0) + F_p(p_0,q_0) (p-p_0) + F_q(p_0,q_0) (q-q_0),
\end{eqnarray}
which coincides with expression (\ref{Fex1}) if $\alpha_0= 
F(p_0,q_0) - F_p(p_0,q_0) p_0- F_q(p_0,q_0) q_0$, $\alpha_1 =F_p(p_0,q_0) $,
$\alpha_2 =
F_q(p_0,q_0) $. Consequently, function (\ref{Fex1}) is responsible for the
local solvability of 
PDE (\ref{intr}).

{\bf Example 2.} Next, we suggest the following function:
\begin{eqnarray}\label{Fex2}
F=e^{\alpha p} + c_2 e^{\beta p + c_1 q} + c_3 e^{\frac{\alpha c_1
q}{\alpha-\beta}},
\end{eqnarray}
where $\alpha$, $\beta$, $c_i$, $i=1,2,3$, are arbitrary parameters.
In this case
\begin{eqnarray}
&&
{\cal{M}} = a_1 F + a_2 F^2 ,\\\nonumber
&&
a_1=-\left(\frac{\alpha c_1}{\alpha-\beta} 
\Big((\alpha-\beta) c_2  e^{\beta p+  c_1 q} +\alpha  
c_3 e^{\frac{\alpha c_1 q}{\alpha-\beta}}\Big)\right)^2
,\\\nonumber
&&
a_2= (\alpha c_1)^2\left(
 c_2 e^{\beta p+  c_1 q} +\frac{\alpha^2
c_3}{(\alpha-\beta)^2}    e^{\frac{\alpha c_1 q}{\alpha-\beta}}\right).
\end{eqnarray}

{\bf Example 3.} Another  example of solution to eq.(\ref{gF}) is following:
\begin{eqnarray}\label{Fex3}
F=e^{2 p} +e^p g(q) + e^{q c_1} c_2 ,
\end{eqnarray}
where $c_1$ and $c_2$ are arbitrary constants and 
$g(w)$ satisfies the following second order ODE:
\begin{eqnarray}
g'' + \frac{1}{4 e^{q c_1} c_2 -g^2}\Big(g (g')^2  - e^{q c_1} c_1 c_2 
(4  g'-c_1 g )\Big) =0.
\end{eqnarray}
It may be written in terms of variable $s=e^{q c_1}$ and function 
$h(s)=g\left(\frac{\ln(s)}{c_1}\right)$ as
\begin{eqnarray}
(s c_1)^2 \left(
h'' + \frac{ h ( (h')^2 s - h h' +c_2)}{s(4 s c_2 - h^2)}
\right) =0.
\end{eqnarray}
In this case
\begin{eqnarray}
&&
{\cal{M}}= (a_1   + a_2 e^p) F  + (a_3  + a_4 e^p) F^2,\\\nonumber
&&
a_1=- d^{-1} 4 c_2 e^{c_1 q}(2 c_1 c_2 e^{c_1 q} - g g')^2,\\\nonumber
&&
a_2= -8 e^{2 c_1 q} c_1c_2^2 ( 2 g' +c_1 g) -e^{c_1 q} c_2 g 
(c_1^2 g^2 - 16 c_1 g g' -4 (g')^2)-
4 g^3 (g')^2,\\\nonumber
&&
a_3=d^{-1} 4 (2 c_1 c_2 e^{c_1 q} - g g')^2 ,\\\nonumber
&&
a_4= d^{-1} 4 c_1 c_2 e^{c_1 q} ( c_1 g - 4 g') + 4 g (g')^2
,\\\nonumber
&&
d=4 c_2 e^{c_1 q} - g^2 .
\end{eqnarray}

%%%%%%%%%%%%%%%%%%%%%%%%
\section{Spectral equation associated with nonlinear PDE (\ref{uw})}
\label{Section:spectral}

We may enrich the solution space  adding   arbitrary  parameters 
(which might be called spectral parameters) in the function $F$. These parameters may be
introduced     into the function 
 ${\cal{M}}$ in the rhs of eq.(\ref{gF}). Such parameters appear in 
 solution
(\ref{Fex2}) (parameters $\alpha$, $\beta$, $c_i$, $i=1,2,3$) and in solution
(\ref{Fex3})
 (parameters $c_1$, $c_2$). In addition, two arbitrary functions of single variable 
(say, $C_i(q)$, $i=1,2$) appear  in the course of integration of the second order two-dimensional 
nonlinear PDE (\ref{gF}). 
 Let us denote the set of all arbitrary parameters 
by the vector parameter $\lambda$ and use it as follows.

Instead of the variable $w$ (which is a function of $u$ and its $x$-derivatives) 
in the list of parameters of the function $F$ we use
another 
variable $W(\lambda,C_1,C_2)$
 related with $w$ by the integral 
\begin{eqnarray}\label{Wu}
w(u,u_x,u_{xx},\dots)=\int DC_1 DC_2 \int d\Omega(\lambda) W(\lambda,C_1,C_2),
\end{eqnarray}
where $\Omega$ is some measure in the space of the vector 
parameter $\lambda$ and $\int DC_1 DC_2$ means the functional integration with respect 
to the functions $C_i$, $i=1,2$.
 This equation must be considered as a PDE for the function $u$, 
 which is independent on $\lambda$.
Now we have to replace equations 
(\ref{ex1F}) and (\ref{F0}) with the following ones,
respectively
\begin{eqnarray}\label{FW1}
F(\xi,W(\lambda,C_1,C_2),\lambda) =0,\\\label{FW2}
F(u,W(\lambda,C_1,C_2),\lambda) =0.
\end{eqnarray}
These equations must be solved for $W$. 
Applying the same algorithm as in Secs.\ref{Section:first} and
\ref{Section:second}
we derive equations
\begin{eqnarray}\label{FWpsi}
F_\xi t E(u) + F_W E(W) =0,\\
\label{2FWpsi}
F_u E(u) + F_W E(W) =0
\end{eqnarray}
instead of eqs.(\ref{Fuw}) and (\ref{Fuwex2}) respectively.
Here 
\begin{eqnarray}\label{WE}
E(W) = W_t + Q W_{x_1}.
\end{eqnarray}
Now we divide eqs.(\ref{FWpsi}) and (\ref{2FWpsi}) 
by $F_W$ and apply the operator 
$\int DC_1 DC_2 \int d\Omega(\lambda)$.
In virtue of eqs.(\ref{Wu}) and (\ref{uw2}) we obtain
\begin{eqnarray}\label{WFpsi2}
\left(\int DC_1 DC_2\int d\Omega(\lambda) \frac{F_{\xi} t}{F_W} + Q\right) \psi
=0,\\\label{2WFpsi2}
\left(\int DC_1 DC_2\int d\Omega(\lambda) \frac{F_u}{F_W} + Q\right) \psi =0.
\end{eqnarray} 
Note that $Q$ and $\psi$ are different in eqs.(\ref{WFpsi2}) and (\ref{2WFpsi2}), 
see eqs.(\ref{Q},\ref{psi}) and (\ref{QQ},\ref{psi2}) respectively.
Now
we may use the zero 
boundary conditions discussed in Theorems 1 and 2. Finally, solving eq.(\ref{Wu}) 
for $u$ with the above boundary conditions we obtain a solution to the original $(M+1)$-dimensional PDE. 

Now we turn to eqs.(\ref{FWpsi}) and (\ref{2FWpsi}). Since the constructed $u$ is a solution 
to the original nonlinear PDE,   the first terms in  eqs.(\ref{FWpsi}) and (\ref{2FWpsi})  vanish. 
Consequently, since $F_W\neq 0$, we derive a linear
PDE for the spectral function $W(\lambda)$:
\begin{eqnarray}\label{SPeq}
W_t(\lambda,C_1,C_2) + Q W_{x_1}(\lambda,C_1,C_2) =0.
\end{eqnarray} 
This equation can be considered as a spectral equation for the 
nonlinear PDE (\ref{intr}). 
Emphasize that, unlike the ISTM,
we have only one spectral problem associated
with the given nonlinear PDE. Consequently, 
PDE (\ref{uw}) may not be viewed as the compatibility condition for some
overdetermined system of linear PDEs.

\subsection{On the richness of the solution space to the  $(M+1)$-dimensional
nonlinear PDE (\ref{intr})}
The richness of solution space to  general  $(M+1)$-dimensional 
nonlinear PDE (\ref{intr})
is defined by two sets of arbitrary functions. First set 
consists of two arbitrary functions of single variable
appearing in the general solution of   two-dimensional second order PDE
(\ref{gF}) with the given function ${\cal{M}}$. Moreover,  the 
function ${\cal{M}}$ is an arbitrary function 
of three arguments (restricted by the only condition (\ref{calM})) which, 
in addition, might  arbitrarily depend on the  spectral parameter $\lambda$. 
The second set appears in the general solution to either 
eq.(\ref{ex1F}), or (\ref{F0}), or (\ref{Wu}),  which are the $M$-dimensional 
(in general) nonlinear PDEs due to the fact that 
 $w$ is an $M$-dimensional differential expression of $u$. Thus, its general 
 solution depends on 
 the arbitrary functions of $M$ variables: $(M-1)$ variables from the list $x$ 
 supplemented by the variable $t$. However, these arbitrary functions satisfy 
 the  zero boundary conditions for the function $\psi = u_t + w_{x_1}$,  
 which are discussed in 
 Theorems 1 and 2 (see Secs.\ref{Section:first} and \ref{Section:second}
 respectively). This fact    imposes additional constraints on the above arbitrary
functions.
  The unresolved problem is whether the 
 combination of these two sets of arbitrary functions together with arbitrary function ${\cal{M}}$ 
 and spectral parameter may lead to  the 
 complete integrability of the nonlinear PDE (\ref{ex2ut1}).
This problem 
 is left for further study. The role of functional integration in equation (\ref{Wu}) 
 should be also clarified.

 \section{Solvability of eqs.(\ref{ex1F}) and (\ref{F0})}
 \label{Section:F}
 After the function $F(p,q)$ is constructed, we have to solve $M$-dimensional
PDE, 
 either  
 (\ref{ex1F}) or  (\ref{F0}), for the function $u$.
 Thus we deal with the integrability of these equations.
As a simple case  we consider eq.(\ref{F0}) of Sec.\ref{Section:second} with
$F(u,w) = u + w$:
\begin{eqnarray}\label{comF02}
u+w=0.
\end{eqnarray}
This equation may be integrated, for instance,  in the following cases.

\subsection{Eq.(\ref{comF02})  is integrable by the inverse spectral transform
method: 
 short-pulse equation}
First, we consider the case when  eq.(\ref{comF02}) is
integrable by ISTM. For instance, let 
eq.(\ref{comF02}) be a short-pulse equation (SPE) \cite{SW,SS}
\begin{eqnarray}\label{SPE}
u_{x_2x_3} + (u^3)_{x_2x_2} + u =0,
\end{eqnarray}
i.e. 
\begin{eqnarray}
w=u_{x_2x_3} + (u^3)_{x_2x_2}.
\end{eqnarray}
In this case eq.(\ref{ex2ut1}) reads
\begin{eqnarray}\label{SPEu}
u_t + (u_{x_2x_3} + (u^3)_{x_2x_2})_{x_1} =0
\end{eqnarray}
and may be considered as an additional commuting flow to SPE (\ref{SPE}).
Remember, that not any solution of SPE satisfies eq.(\ref{SPEu}), but only those
that 
satisfy the zero boundary condition for the function
\begin{eqnarray}
\psi = u_t + (u_{x_2x_3} + (u^3)_{x_2x_2})_{x_1}
\end{eqnarray}
as a solution to a 2-dimensional PDE (\ref{2Fpsi}) where $F_u=F_w=1$ 
and the differential operator $Q$ is defined as 
$Q=6 (u_{x_2}^2+u u_{x_2x_2}) + 12 u u_{x_2} \partial_{x_2} + 
3 u^2 \partial_{x_2x_2} + \partial_{x_2x_3}$.

\subsection{Linearizable eq.(\ref{comF02})}

Now we consider the linearizable  equation (\ref{comF02}). For instance, 
if  $w=-\ln(u_{x_3}+ u_{x_2x_2}- u_{x_2}^2)$, then eq.(\ref{comF02}) is
linearizable by the 
substitution $u = -\ln \varphi $  yielding 
\begin{eqnarray}
\ln(-\varphi_{x_3}-\varphi_{x_2x_2}) =0\;\;\Rightarrow
\;\;\varphi_{x_3}+\varphi_{x_2x_2} +1=0.
\end{eqnarray}
The corresponding PDE (\ref{ex2ut1}) reads
\begin{eqnarray}
u_t - \ln_{x_1}(u_{x_3}+ u_{x_2x_2}- u_{x_2}^2) =0.
\end{eqnarray}
Therewith the function $\psi = u_t - \ln_{x_1}(u_{x_3}+ u_{x_2x_2}- u_{x_2}^2)$ 
must satisfy the zero boundary condition for the PDE (\ref{2Fpsi}) with 
$F_u=F_w=1$ and $\displaystyle Q=-\frac{\partial_{x_3}+
\partial_{x_2x_2}-
2 u_{x_2} \partial_{x_2}}{u_{x_3}+ u_{x_2x_2}- u_{x_2}^2}$.

\subsection{Eq. (\ref{comF02}) reducible to ODE}
\label{Section:ODE}
In this subsection we consider the case when eq.(\ref{comF02}) itself
can be represented in form (\ref{intr}):
\begin{eqnarray}\label{w1}
u+w \equiv u_{x_1} + w^{(1)}_{x_2} (u,u_{y_1},u_{y_1y_1},\dots) =0
\end{eqnarray}
with some function $w^{(1)}$ of $u$ and its derivatives with respect to $x_i$, $i=2,3,
\dots, M$. 
Hereafter in this section  we use the lists $y_i$, $i=1,2,\dots,M-1$,
of variables $x_i$ defined as
$y_i=\{x_{i+1},\dots,x_M\}$, and functions $w^{(i)}(u,u_{y_i},u_{y_iy_i},\dots)$ 
depending on $u$ and its derivatives with respect to the variables $x_{i+1}$,
$x_{i+2}$, $\dots$, $x_{M}$.
Thus, eq.(\ref{w1}) has solutions defined by equation of form (\ref{comF02}):
\begin{eqnarray}
\label{wF1}
u+w^{(1)} =0.
\end{eqnarray}
If, in turn, eq.(\ref{wF1}) is representable in form (\ref{intr}),
\begin{eqnarray}
u+w^{(1)}\equiv u_{x_2} + w^{(2)}_{x_3} (u,u_{y_2},u_{y_2y_2},\dots) =0,
\end{eqnarray}
then it possesses solutions defined by equation of form (\ref{comF02}):
\begin{eqnarray}
\label{wF2}
u+w^{(2)} =0
\end{eqnarray}
and so on. If this process may be continued, then, on the $i$th step, we have the equation
\begin{eqnarray}
u+w^{(i-1)}\equiv u_{x_i} + w^{(i)}_{x_{i+1}} (u,u_{y_i},u_{y_iy_i},\dots) =0,
\end{eqnarray}
which possesses solutions defined by equation of form (\ref{comF02}):
\begin{eqnarray}
\label{wFi}
u+w^{(i)} =0.
\end{eqnarray}
On the last step $y_{M-1} =x_M$, so that we end up with ODE:
\begin{eqnarray}
u+w^{(M-1)}(u,u_{x_M}, u_{x_M x_M},\dots,) =0. 
\end{eqnarray}
On each step we have to provide a proper zero boundary conditions 
for the function $\psi^{(i)}=u_{x_i} + w^{(i)}_{x_{i+1}})$ as  a  solution to 
the linear  differential  
equation $(1 + Q^{(i)})\psi^{(i)}=0$ with the differential operator $Q^{(i)}=w^{(i)}_u  +
\sum_{j=1}^M w^{(i)}_{ u_{x_j}} \partial_{ x_j} +
 \sum_{j_1,j_2=1}^M w^{(i)}_{ u_{x_{j_1}x_{j_2}}} \partial_{ x_{j_1}x_{j_2}} 
 + \dots$.
 Remember that PDE (\ref{w1}) itself is supplemented by the boundary condition for the function 
 $\psi = u_t +w_{x_1}$ as a solution to PDE $(1+Q)\psi=0$ with $Q$ 
 given by expression (\ref{QQ}).
The simplest equation of this form is following
\begin{eqnarray}
u_{t} + (u_{x_1} -u+ f_{x_2}(u) )_{x_1} \equiv u_t -u_{x_1} + u_{x_1x_1} + 
f_{x_2 x_1}(u) =0,
\end{eqnarray}
where $f$ is an arbitrary function of $u$.

\section{Conclusions}
\label{Section:conclusions}

We represent an algorithm reducing a  
large  class of ($M+1$)-dimensional nonlinear
evolutionary 
 PDE having the form of one-dimensional flow  (\ref{intr}) to a family of  $M$-dimensional nonlinear PDEs (\ref{F0}) 
 (or (\ref{ex1F}) if $w$ is in the form (\ref{w})). The compatibility of ($M+1$)-dimensional 
 and $M$-dimensional PDEs leads to 
the two-dimensional second order PDE (\ref{gF})  for the function  $F$.  
It is remarkable, that this PDE holds for any
($M+1$)-dimensional PDE of the form (\ref{intr}). Moreover, a vector spectral parameter 
may be introduced into the PDE for $F$. 
The presence of this parameter leads to the single 
linear spectral equation associated with the original 
($M+1)$-dimensional nonlinear PDE.
However, there is no second spectral equation 
which would compose the Lax representation, like in the case of the ISTM.
 The solvability of  the  equation for $F$
 is the first nontrivial step in construction of particular  solutions since this equation 
 is nonlinear one, which will be studied in a different paper. Here we  represent several particular solutions
to this equation in Sec.\ref{Section:solutionsF}. Another problem is solvability of the 
 $M$-dimensional PDE  $F=0$ for the function $u$, which 
 involves  the function $w$ from eq.(\ref{intr}) 
 and must be studied in each particular case. 
   In some cases, we may reduce  the ($M+1$)-dimensional 
PDEs to ODEs, which, sometimes, might be explicitly solved, as is 
demonstrated in the examples of Secs.\ref{Section:Example1} 
and \ref{Section:Example2}, see also Sec.\ref{Section:ODE}.

   It is obvious that 
there is a large manifold of the nonlinear PDEs from the derived class 
that have a physical application. We may refer to the nonlinear PDEs considered
in 
ref.\cite{Z_KdV}, which concern  our case as well. 
A deformation of one of these equations is considered in Sec.\ref{Section:Example1}.

The author thanks Prof. L.V.Bogdanov for usefull discussion. 
This work is supported by 
the Program for Support of Leading Scientific Schools 
(grant No. 6170.2012.2).

%%%%%%%%%%%%%%%%%%%%%%%%%%%%%%%%%%
%%%%%%%%%%%%%%%%%%%%%%%%%%%%%%%%%

\end{document}